\documentstyle[12pt,epsfig]{article}
\newcommand{\nn}{\nonumber}

\newcommand{\tr}{{\rm Tr}}
\begin{document}

\pagestyle{empty}
\begin{flushright}
CERN-TH/96-122 \\
ROME1-96/1147\\
\end{flushright}
\vskip 0.5cm
\centerline{\bf HYBRID MOLECULAR DYNAMICS FOR LATTICE SUPERSYMMETRY}
\vskip 0.8cm
\centerline{\bf A. Donini$^{a,b}$ and M. Guagnelli$^a$}
\centerline{$(a)$ Theory Division, CERN, 1211 Geneva 23, Switzerland,}
\centerline{$(b)$ Dip. di Fisica,
Universit\`a degli Studi di Roma ``La Sapienza'' and}
\centerline{INFN, Sezione di Roma, P.le A. Moro 2, 00185 Rome, Italy }

\vskip 3cm
\begin{abstract}
We present the first results obtained with a Hybrid Molecular Dynamics
algorithm applied to an $N=1$ SU(2) Super-Yang--Mills on the lattice. We 
derive the Hamilton equations of motion for the system with Wilson gluinos
and present preliminary results
on small lattices. 
\end{abstract}
\vskip 5cm

\begin{flushleft}
CERN-TH/96-122 \\
May 1996
\end{flushleft}
\vskip 0.5 cm

\newpage
\pagestyle{plain}
\setcounter{page}{1}

\section{Introduction}
\label{sec:intro}

The supersymmetric extension of the Standard Model is now considered the
most promising solution to the gauge hierarchy problem; this problem
arises whenever the SM itself is considered as the low-energy
effective action of a more general theory including gravity. 
Moreover, it has recently been found 
that exact non-perturbative solutions of supersymmetric Yang--Mills theories
can be found due to the peculiar properties of extended $N=2$ supersymmetry. 
It is then interesting to begin the 
exploration of the non-perturbative aspects of supersymmetric theories by
using their lattice formulation, in order to achieve a better understanding of
the confinement and chiral symmetry breaking phenomena.
The lattice formulation of a supersymmetric theory seems to be
impossible since the discretization of the space--time explicitly breaks 
supersymmetry. Nevertheless, it has been shown \cite{CUVE} that it is 
not necessary to put on the lattice an explicitly supersymmetric theory: 
much as in the case of chiral symmetry, the main point is that the
continuum limit of the theory should be supersymmetric.
Thus, it is possible to simulate a non-supersymmetric theory  with the same 
field content as the continuum one, with the requirement that in the 
$a \rightarrow 0$ limit the supersymmetric theory is recovered.
It has been shown (for $N=1$ Yang--Mills in \cite{CUVE} and for $N=2$ in 
\cite{MONT2}) that the supersymmetric limit actually is strongly related to 
the chiral limit itself. In this way, in order to study $N=1$ 
Super-Yang--Mills (SYM) it is only needed to introduce gluons and massive 
gluinos (fermion fields in the same representation as the gauge group
of the gluons)
and then recover the supersymmetric limit going to the $m_g = 0$ chiral one.
It is evident from what was said above that in this theory it is fundamental to
maintain an exact balance between bosonic and fermionic degrees of freedom.
For this reason, an unquenched simulation should be performed, 
keeping also gluino loops. 
A recent paper \cite{MONT1} presented a first attempt to do
an $N=1$ SU(2) Yang--Mills simulation on small lattices using the L\"uscher
local bosonic algorithm. In order to perform a cross-check of the results, we
have decided to adopt a different approach, namely the Hybrid Molecular 
Dynamics (HMD) algorithm of \cite{GOTT}. 

In section \ref{sec:action} we introduce the Wilson action for SYM
theory and the Hamilton equations of motion; in section \ref{sec:res} we
present our preliminary results, obtained on a small lattice, and
finally in section \ref{sec:concl} we draw some conclusions
discussing the feasibility of these kind of simulations and giving an
outline of our future projects. 

\section{Supersymmetric Wilson action on the lattice}
\label{sec:action}

In this section the basic formalism, the notation  and the main
properties of the $N=1$ supersymmetric extension of the SU(2)
Yang--Mills theory are introduced. 
The Hamilton equations for the Molecular Dynamics algorithm
are also given. 
 
We recall that the parameter $N$ refers to the number of
anti-commuting generators that are introduced in order to make a
theory supersymmetric. The easiest supersymmetric extension of a
non-Abelian gauge theory is $N=1$ SU(2) Yang--Mills, and so we will
take this theory as our starting point for a study of the
non-perturbative aspects of supersymmetric theories with the lattice approach.
The $N=1$ Yang--Mills action contains two different pieces: the
first is the usual Wilson action for the gauge bosons and the second
represents the action term for the gauge fermions, the gluinos.  We
have adopted as fermion action the Wilson formulation in order to avoid the
doubling  of fermionic flavours. The full SYM  action is
\begin{equation}
S_{SYM} = S_g + S_f,
\end{equation}
where the standard Wilson action for the SU(2) gauge field is
\begin{equation}
S_g = - \frac{\beta}{2} \sum_{\Box} \tr ( U_{\Box} ).
\end{equation}
The fermionic action should describe Majorana spinors in the adjoint
representation of the gauge group. It is possible \cite{MONT1}
to derive an expression
formally identical to the usual Wilson action for the quark in QCD, replacing 
the link variables $U_{\mu}$ that are $N_c \times N_c$ matrices (where
$N_c$ is the rank of the gauge group) with new fields in the adjoint 
representation $V_{\mu}$, using $ (N_c^2 - 1) \times (N_c^2 - 1)$ 
matrices. We introduce the Majorana fermions as 
\begin{equation}
  \label{lambda}
  \lambda(x) = T^{a}\lambda_{a}(x)
\end{equation}
with $T^{a}$ the gauge group generators, satisfying the trace condition
\begin{equation}
  \label{trace}
  \tr[T^{a}T^{b}] = \frac{1}{2}\delta^{ab}.
\end{equation}
For SU(2) we can take $T^{a} = \frac{1}{2}\sigma^{a}$, the Pauli
matrices.
We recall the gauge transformation properties of links and gluinos:
\begin{eqnarray}
  \label{gautra}
  U_{\mu}(x) &\to& \Lambda(x) U_{\mu}(x) \Lambda^\dagger(x + \mu) , \nn \\
  \lambda(x) &\to& \Lambda(x) \lambda (x) \Lambda^\dagger(x) .
\end{eqnarray}

The fermionic action, in the Wilson approach, is then given by
\begin{eqnarray}
  \label{s_ferm1}
S_f &=& \frac{1}{2} \sum_x \left \{ \bar \lambda^a(x) \lambda^a (x) 
\right. \nn \\
&-& \left. 
K \sum_{\mu} \left [ \bar \lambda^a (x + \mu) V^{ab \dagger}_{\mu} (x)
(1 + \gamma_{\mu}) \lambda^b (x) + \bar \lambda^a (x) V^{ab}_{\mu}(x)
(1 - \gamma_{\mu})\lambda^b(x + \mu) \right ] \right \},
\end{eqnarray}
where the adjoint link $V_{\mu}$ is given by
\begin{equation}
  \label{link}
  V^{ab}_{\mu}(x) = \frac{1}{2} \tr [ U^{\dagger}_{\mu}(x) \sigma^a
U_{\mu}(x) \sigma^b ] .
\end{equation}
It is easy to verify that $V_{\mu}$ belongs to O(3), and so $V^{\dagger}_{\mu}
= V^T_{\mu}$.
In the continuum limit, the fermionic action (\ref{s_ferm1}) gives 
the interaction Lagrangian
\begin{equation}
{\cal L}_I = \frac{1}{2} f^{a b c} \bar \lambda^a A^b_{\mu}
\gamma_{\mu} \lambda^c.
\end{equation}
The action can be rewritten in the following way:
\begin{equation}
  \label{fermion_action}
  S_{f} = -\frac{1}{4} \tr \left [ \ln ( M^{\dagger} M ) \right ],
\end{equation}
where $M$ is the fermionic matrix.
The factor $\frac{1}{4}$ in eq. (\ref{fermion_action}) contains a
$\frac{1}{2}$ due to the square root of $\det M^\dagger M$ and another
$\frac{1}{2}$ that takes into account the Majorana nature of the gluinos.
A detailed description of how the gluino propagator and the $n$-point
correlation functions can be obtained in the path integral formalism on
the lattice can be found in \cite{MONT1}.

In order to simulate this theory, we have adopted the Hybrid Molecular
Dynamics approach introduced in \cite{GOTT}. Since the
Super-Yang--Mills theory we want to reproduce contains just one gluino 
flavour, we are forced to use the so-called $R$-algorithm.
For this reason we cannot adopt the popular Hybrid Monte Carlo approach,
which could be used only with a multiple of 4 Majorana flavours.
As usual we introduce the Hermitian field $P_{\mu}$ conjugate to the gauge 
field in order to write the Hamiltonian of the system as
\begin{equation}
  \label{hami}
  {\cal H} = \frac{1}{2} \sum_{x,\mu} \tr P^2_{\mu}(x) + S_g + S_f .
\end{equation}
The conjugate momenta $P_{\mu}$ of the link variables are generated at
the beginning of every Molecular Dynamics trajectory as complex Gaussian
numbers distributed as $\exp (- \frac{1}{2} \tr P^2_{\mu} )$. 

Following \cite{GOTT} we introduce a noisy estimator for $(M^\dagger
M)^{-1}$ in the form $\chi \chi^\star$, where $\chi$ is the solution of
the matrix equation
\begin{equation}
 \label{mateq}
 M^\dagger M \chi = \phi
\end{equation}
and $\phi = M^\dagger R$. In this expression $R$ is a vector of random
Gaussian complex numbers with the same indices as the original gluino
field, distributed as $\exp (- R^\star R)$.

The Hamilton equations for the system are
\begin{eqnarray}
  \label{hameq1}
  \dot{U}_{\mu}(x) &=& i P_{\mu}(x) U_{\mu}(x) , \nn \\ 
  \dot{P}_{\mu}(x) &=& \frac{i}{2}\beta{\cal T}[G_\mu(x)] 
                     - \frac{i}{2}K{\cal T}[F_\mu(x)] ,
\end{eqnarray}
where the derivative is taken with respect to the Molecular Dynamics
``time'' and ${\cal T}$ is the projection operator on traceless anti-Hermitian
matrices:
\begin{equation}
  \label{traceless}
  {\cal T}[B] = \frac{1}{2} \left [ (B - B^\dagger) - 
                    \frac{1}{N_c}\tr(B - B^\dagger) \right ] .
\end{equation}

The evolution of the conjugate momenta $P_{\mu}$ fields is driven by the
gauge ``force''
\begin{equation}
  \label{staple}
  G_{\mu}(x) = U_{\mu}(x) \sum_{\nu\ne\mu} \left [ 
       U_{\nu}(x+\mu)U^\dagger_{\mu}(x+\nu)U^\dagger_{\nu}(x) +
       U^\dagger_{\nu}(x+\mu-\nu)U^\dagger_{\mu}(x-\nu)U_{\nu}(x-\nu)
       \right ]
\end{equation}
and by the fermion ``force''
\begin{equation}
  \label{ferfor}
  F_{\mu}(x) = A_{\mu}^{ab}(x) {\rm Re} [ B_{\mu}^{ba}(x)] ,
\end{equation}
where
\begin{eqnarray}
  \label{aandb}
  A_{\mu}^{ab}(x) &=& \sigma^a U^\dagger_{\mu}(x)\sigma^b U_{\mu}(x) , \nn \\
  B_{\mu}^{ba}(x) &=& \tr \left [
                    (1 - \gamma_{\mu})Y^b(x)\chi^{a\star}(x+\mu) +
                    (1 + \gamma_{\mu})\chi^b(x)Y^{a\star}(x+\mu)
                   \right ]_{\rm Dirac} 
\end{eqnarray}
and $Y = M\chi$.

As in \cite{MONT1}, we assume that the sign of the Pfaffian operator (the
square root of the fermion determinant) remains positive during the
evolution of the system in the Molecular Dynamics ``time''. In principle, 
we could take this phase into account by inserting it in the definition
of the observables.

\section{Preliminary results}  
\label{sec:res}

In this section we present the first results we obtained with the Hybrid
Molecular Dynamics algorithm for the Super-Yang--Mills SU(2) $N=1$ theory.

We have performed a test of the method using the same parameters as
reported in \cite{MONT1}, namely $\beta = 2.0$ and $K = 0.150$. 
We have run the program on a lattice with the same size ($4^3 \times
8$), obtaining compatible results for the plaquette, the absolute
value of the Polyakov loop and the $ 2 \times 2 $ Creutz ratio,
defined in terms of $ k \times l $ Wilson loops $W(k,l)$ as

\begin{equation}
  \chi_2 = \frac{W(2,2) W(1,1)}{W(2,1) W(1,2)} .
\end{equation}

In all the runs the 
time--length of the Molecular Dynamics trajectories between 
different measurements was kept fixed at 0.5. This implies a number of
{\em leap-frog} steps $N_{step} = 5, 10, 20$ for the three simulations
performed.
Our statistics is of the order of $N_{m} = 10000$ measures
for the full theory runs, and $N_{m} = 5000$ in the pure gauge case.
In this last case, we have used method 1 of \cite{GUPTA}.
With a short heat-bath run we obtain for the SU(2) pure gauge plaquette
at $\beta= 2.0$ the value $\Box_{PG} = 0.5008(3)$.

\begin{figure}[htbp]
  \begin{center}
    \leavevmode
    \epsfig{figure=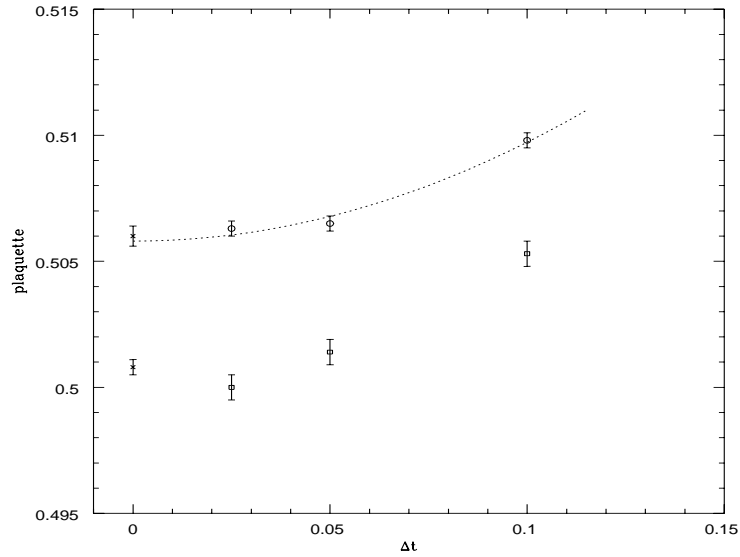,height=8cm,width=10cm,angle=0}
    \caption{Average plaquette versus $\Delta t$ in the pure gauge case 
             (squares) and within the full theory (circles). The upper 
             cross at $\Delta t = 0$ is reported by Montvay, while
             the lower one is the heat bath pure gauge result. The
             dotted line is the extrapolation to $\Delta t = 0$. }
    \label{fig1}
  \end{center}
\end{figure}

It is important to stress that the algorithm used introduces an intrinsic
${\cal O}(\Delta t^2)$ error, where $\Delta t$ is the Molecular Dynamics time
step. It is then necessary to extrapolate to the $\Delta t = 0$ limit in 
order to obtain physical results. Furthermore, the approach of the plaquette
to the physical value is expected to be from above. Both behaviours,
namely the $\Delta t^2$ dependence and the approach from above can be
seen in fig. \ref{fig1} where the expectation value of the plaquette
is presented as a function of $\Delta t$.

\begin{table}[htbp]
  \begin{center}
    \leavevmode
    \begin{tabular}[]{|c|c|c|c|c|}
\hline
 $\Delta t$ & $\Box$      &    $|P|$    &  $\chi_2$    & $N_m$   \\
\hline\hline
  0.100     & $0.5098(3)$ & $0.0505(4)$ & $0.5555(12)$ & $10000$ \\
  0.050     & $0.5065(3)$ & $0.0500(4)$ & $0.5559(13)$ & $8000$  \\
  0.025     & $0.5063(3)$ &     -       &     -        & $10000$ \\
\hline          
    \end{tabular}
    \caption{Average observables for
 $\beta=2.0$, $K=0.150$, $V=4^3\times 8$ at different values of $\Delta
 t$. We show the plaquette, the absolute value of the Polyakov loop
 and the $2 \times 2$ Creutz ratio. $N_m$ is the number of
 measurements for each run.}
    \label{tab1}
  \end{center}
\end{table}

We present in table \ref{tab1} the results obtained for the three
observables introduced above. The results are fully compatible with
those reported in \cite{MONT1}: for example the value of the plaquette
extrapolated to $\Delta t = 0$ is $\Box = 0.5058(4)$.
We have no data for $|P|$ and $\chi_2$ at $\Delta t = 0.025$, but the
values obtained at larger $\Delta t$ are quite flat and already
compatible with \cite{MONT1}.

\begin{figure}[htbp]
  \begin{center}
    \leavevmode
    \epsfig{figure=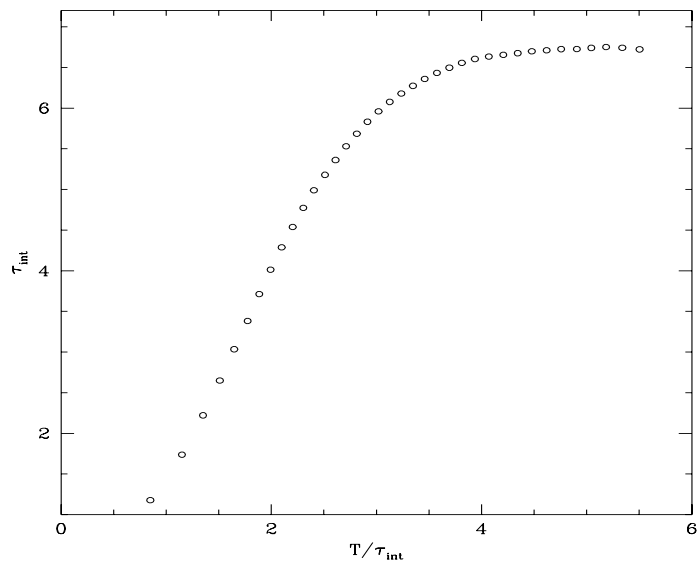,height=8cm,width=10cm,angle=0}
    \caption{Integrated autocorrelation time versus $T/\tau_{\rm int}$
             for the plaquette at $\Delta t = 0.025$.}
    \label{fig2}
  \end{center}
\end{figure}

On the same observables, we have studied the autocorrelation times at
different time steps, with the windowing method \cite{SOKAL}.
We build the connected autocorrelation function $C(t)$ and from this
an estimate for the integrated autocorrelation time
\begin{equation}
  \label{tauint}
  \tau_{\rm int}(T) = \frac{1}{2} + \sum_{t=1}^{T}\frac{C(t)}{C(0)} .
\end{equation}
When the ratio $T/\tau_{\rm int}(T)$ is around 4 we should observe a
plateau in the value of $\tau_{\rm int}(T)$: this is exactly what
happens, as can be seen in fig. \ref{fig2} in a particular case.
Results for integrated autocorrelation times are presented in table
\ref{tab2}, together with a rough estimate of the errors.

\begin{table}[htbp]
  \begin{center}
    \leavevmode
    \begin{tabular}[]{|c|c|c|c|}
\hline
 $\Delta t$ & $\tau_{\Box}$ & $\tau_{|P|}$ &  $\tau_{\chi_2}$ \\
\hline\hline
  0.100     & $7.5(5)$      & $1.0(2)$     & $2.8(3)$         \\
  0.050     & $6.8(5)$      & $1.0(2)$     & $2.7(3)$         \\
  0.025     & $6.7(5)$      &    -         &    -             \\
\hline          
    \end{tabular}
    \caption{
       Autocorrelation times for the different observables at
       different values of $\Delta t$.}
    \label{tab2}
  \end{center}
\end{table}

Finally, we present a rough quantitative comparison of the relative
speed of the two algorithms 
used for these simulations (namely, HMD and the L\"uscher
local bosonic one adopted in \cite{MONT1}); an extensive study
of the scaling of the autocorrelation times with the different parameters
relevant to the simulations has yet to be performed and, moreover,
it is likely that further improvements can be implemented.
Since the most time-consuming part of both algorithms is the
application of the fermion matrix to a vector, we consider the
``effort'' $E$ needed to obtain an independent configuration from a
previous one as the number of times this application is performed,
times the autocorrelation time of a particular observable (for
definiteness, the plaquette). For each simulation, introducing $N_{app}$
as the number of times the fermion matrix is applied per step, we obtain
\begin{equation}
E_{HMD} = N_{step} \times N_{app} \times \tau .
\end{equation}
For this preliminary study we have not yet tried to optimize the
algorithm: in particular, so as to solve the matrix equation (\ref{mateq}),
we used the Stabilized Bi-Conjugate Gradient without preconditioning. 
With this particular inverter and with the {\em leap-frog} scheme used
to integrate the Hamilton equation,
\begin{equation}
N_{app} = 2 + 4 N_{iter},
\end{equation}
where $N_{iter}$ is the average number of iterations required to invert the
fermion matrix. The convergence criterion we used for all our runs was
$ | M^\dagger M  \chi  - \phi |^2 < 10^{-8} | \chi |^2 $, obtaining 
$N_{iter} \simeq 25$. 
For the most time-consuming simulation ($\Delta t = 0.025, 
N_{step} = 20$), $E_{HMD} \simeq 14000$; for the whole set of three
simulations at different $\Delta t$, we obtain $E_{HMD} \simeq 25000$.

For the local bosonic algorithm used in \cite{MONT1}, the ``effort''
is computed as 
\begin{equation}
E_{LB} = N_{field} \times N_{app} \times \tau ,
\end{equation}
where $N_{field}$ is the number of bosonic field used in the
simulation. Using the results reported in \cite{MONT1}, we estimate
(for the two most time-consuming simulations)
$E_{LB} \simeq 90000$ for the one-step algorithm with $24$ bosonic
fields, and $E_{LB} \simeq 25000$ for the two-steps algorithm with
$8$ fields
\footnote{ 
These numbers can be decreased to $44000$ and
$15000$ respectively, by changing the mixture of heat-bath and
over-relaxation sweeps in the update of the bosonic fields \cite{MONT3}.
}.

For a realistic study of the theory with the HMD algorithm, 
it is obvious that the inversion algorithm will have to be accelerated,
for example by means of the standard red--black preconditioning, which 
we are currently implementing. 
Another important improvement of the algorithm can be achieved by using
a higher-order Hamilton equations integration scheme, 
of the kind proposed in \cite{SEXTON}, in order to accelerate the
convergence towards the $\Delta t = 0$ limit.

\section{Conclusions}
\label{sec:concl}

In this paper, following the outline presented in recent papers
\cite{MONT1,MONT2}, we have begun an exploration of the feasibility of
 supersymmetric gauge theories simulations on the lattice.
From a theoretical point of view, it is possible to study these kinds of
theories on the lattice using the same approach adopted for chiral
theories as usual QCD, namely the implementation of a non-chiral
non-supersymmetric theory requiring the recovery of these symmetries in
the continuum limit. 
We have obtained results compatible with those presented by Montvay in
\cite{MONT1} using a completely different algorithm, the Hybrid Molecular 
Dynamics one. From these first simulations it seems possible, with 
the present computer power and the present knowledge of fermionic 
algorithms (such as the local bosonic, the hybrid family or even with
the ``negative flavour number extrapolation'' approach \cite{BERM}),
to afford the study of non-perturbative aspects of supersymmetric
extensions of gauge theories. 

The next step is to begin the investigation of the supersymmetric limit
of the theory, which is expected to be reached jointly with the chiral
limit \cite{CUVE}. In order to do that, we are planning to study the 
asymptotic behaviour of the scalar and the pseudoscalar gluino ``meson''. 
In the OZI approximation it is expected that the pseudoscalar mass
approaches zero when going to the chiral limit, while the scalar should go to 
a non-zero value. 

Finally, we plan to investigate the behaviour of the gluino condensate.
This quantity is expected to be a relevant parameter in the study of
non-perturbative supersymmetry breaking: it is worth while to mention
that a clear comprehension of the supersymmetry-breaking phenomena
is of the uttermost importance if the supersymmetric extension of the 
Standard Model has to be a viable solution to the gauge hierarchy problem.
In order to study this observable, it is necessary to understand its
renormalization properties. It is perhaps important to introduce a
non-perturbative renormalization scheme both on the action 
and on the operators \cite{SOMMER}, which seems to be relevant in the
study of the chiral (supersymmetric) limit of the theory with Wilson fermions.

\section*{Acknowledgements}

We warmly thank R. Sommer and G. Veneziano for helpful 
discussions and suggestions. We also thank A. Brandhuber, I. Montvay,
N. Stella and A. Vladikas for discussions.

\end{document}